\documentclass[showpacs,prb,twocolumn,floats]{revtex4}

\usepackage{graphicx}
\usepackage{amsmath}
\usepackage{amssymb}
\usepackage[colorlinks=true,citecolor=blue,linkcolor=blue]{hyperref}
\usepackage{amsfonts}
\usepackage[latin9]{inputenc}
\usepackage{color}
\usepackage{comment}
\usepackage{epstopdf}

\renewcommand{\vec}[1]{\ensuremath{\boldsymbol{#1}}}

\begin{document}

\title{Comment on ``Creating in-plane pseudomagnetic fields in excess of 1000 T by misoriented stacking in a graphene bilayer"}
\date{\today }
\author{M. Van der Donck}
\email{matthias.vanderdonck@uantwerpen.be}
\affiliation{Department of Physics, University of Antwerp, Groenenborgerlaan 171, B-2020 Antwerp, Belgium}
\author{F. M. Peeters}
\email{francois.peeters@uantwerpen.be}
\affiliation{Department of Physics, University of Antwerp, Groenenborgerlaan 171, B-2020 Antwerp, Belgium}
\author{B. Van Duppen}
\email{ben.vanduppen@uantwerpen.be}
\affiliation{Department of Physics, University of Antwerp, Groenenborgerlaan 171, B-2020 Antwerp, Belgium}


\begin{abstract}
In a recent paper [Phys. Rev. B {\bf 89}, 125418 (2014)], the authors argue that it is possible to map the electronic properties of twisted bilayer graphene to those of bilayer graphene in an in-plane magnetic field. However, their description of the low-energy dynamics of twisted bilayer graphene is restricted to the extended zone scheme and therefore neglects the effects of the superperiodic structure. If the energy spectrum is studied in the supercell Brillouin zone, we find that the comparison with an in-plane magnetic field fails because (i) the energy spectra of the two situations exhibit different symmetries and (ii) the low-energy spectra are very different.
\end{abstract}

\maketitle

He \textit{et al.} relate in Ref. [\onlinecite{pseudo}] the Hamiltonian of electrons in bilayer graphene in the presence of an in-plane magnetic field to the Hamiltonian describing electrons in twisted bilayer graphene. They conclude that because the form of both operators is similar, one can interpret the effect of twisting the two layers with respect to each other as the application of an effective in-plane magnetic field, similar to perpendicular pseudomagnetic fields in strained samples of graphene.\cite{pseudo2} 

Although the analysis of electrons in bilayer graphene with an applied in-plane magnetic field is correct, as shown before,\cite{inplane} their description of electrons in twisted bilayer graphene lacks the effects of the superperiodicity of the total system. These effects are particularly important if one aims to describe electrons near the Dirac points because the superlattice Brillouin zone (sBZ) is much smaller than that of untwisted bilayer graphene and the resulting folding of the band structure leads to the reinstatement of the rotational $C_3$ symmetry around the Dirac points, which is absent in the extended zone scheme, and to hybridization of the Dirac points.

The lack of the inclusion of the superperiodic structure of twisted bilayer graphene is due to the fact that the layer twist is included only after the total bilayer graphene Hamiltonian was approximated around one of the two inequivalent Dirac points in the Brillouin zone, yielding Eq. (4) of Ref. [\onlinecite{pseudo}]. Therefore, this Hamiltonian does not capture the change in the spectrum around the other Dirac points and the coupling between the different Dirac cones of the different layers. In a first approximation, following the authors reasoning, one could, however, write the twisted bilayer graphene Hamiltonian in the layer basis $\Psi = (\Psi_{1},\Psi_{2})$, with $\Psi_{i}$ the 2-spinor describing the atomic orbitals of the $i$-th layer, as follows:
\begin{equation}
\label{eq:twisted_Hamiltonian}
H_4(\vec{k}) = 
\begin{pmatrix}
H_2\left(R(-\frac{\theta_t}{2})\vec{k}\right) & t_{1} I_{A}\\
t_{1} I^{\dag}_{A} & H_2^*\left(R(\frac{\theta_t}{2})\vec{k}\right) 
\end{pmatrix} ~,
\end{equation}
with $t_1$ the interlayer hopping parameter, $R(\theta)$ the rotation matrix over an angle $\theta$, where the matrix $I_{A}=(I_2+\sigma_z)/2$ models the interlayer transitions in a zeroth order approximation and the matrix $H_2(\vec{k})$ is defined as 
\begin{equation}
\label{eq:separate_Hamiltonian}
\begin{split}
&H_2(\vec{k}) =
\begin{pmatrix}
0 & t_0f(\vec{k}) \\
t_0f^*(\vec{k}) & 0
\end{pmatrix}, \\
&f(\vec{k}) = e^{i\frac{k_xa}{\sqrt{3}}}+2e^{-i\frac{k_xa}{2\sqrt{3}}}\cos\left(\frac{k_ya}{2}\right),
\end{split}
\end{equation} 
with $t_0$ the intralayer hopping parameter and $a$ the lattice constant. Eq. (\ref{eq:separate_Hamiltonian}) corresponds to the continuum tight-binding Hamiltonian of a separated graphene layer.\cite{Neto2009} The Hamiltonian in Eq. \eqref{eq:twisted_Hamiltonian} now contains the superperiodic structure of twisted bilayer graphene, in contrast to Eq. (4) of Ref. [\onlinecite{pseudo}].

\begin{figure}[tb]
\centering
\includegraphics[width= 8.5cm]{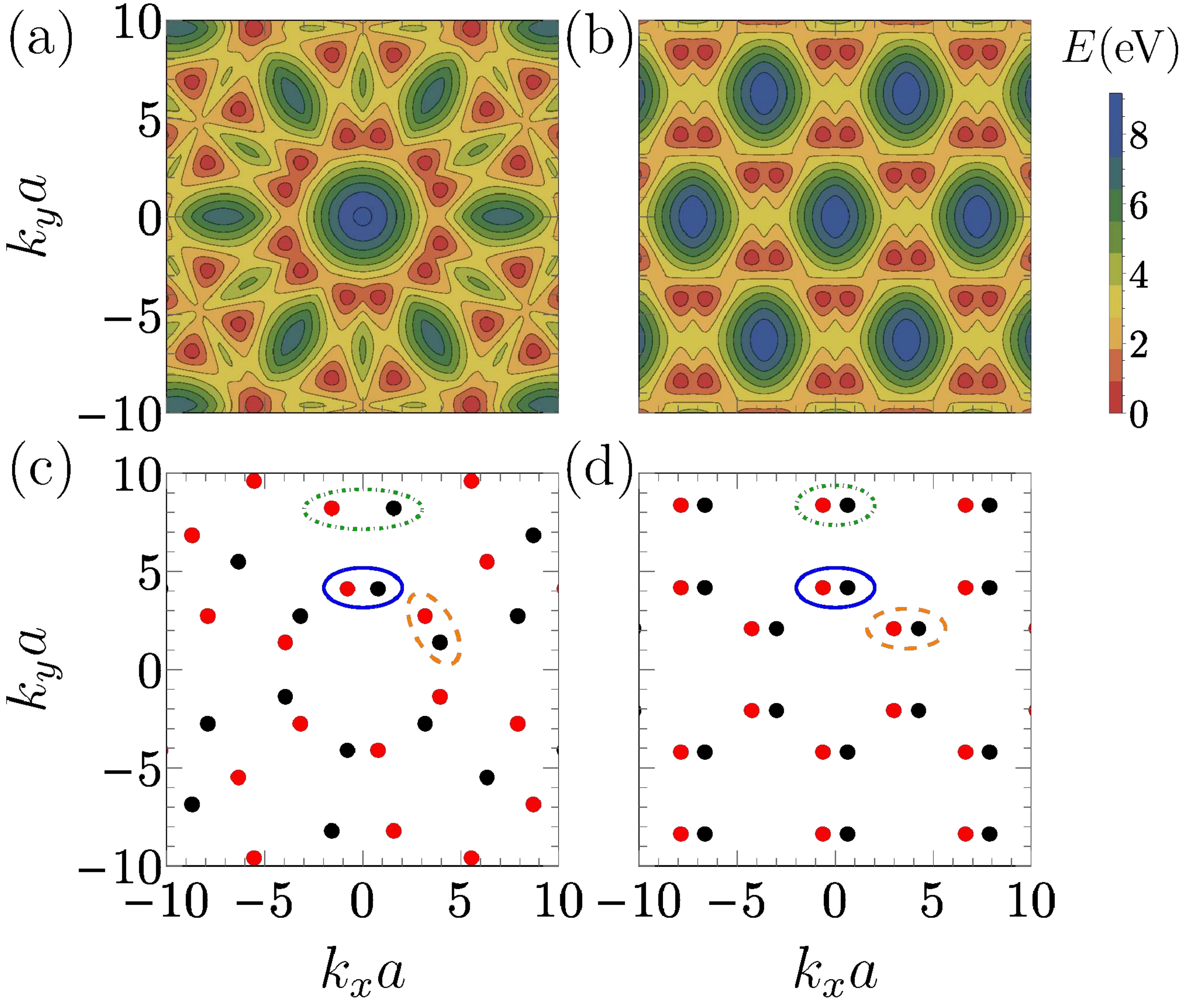}
\caption{(Color online) (a) Lowest conduction band of the energy spectrum of twisted bilayer graphene described by the Hamiltonian (\ref{eq:twisted_Hamiltonian}) for $\theta_t=21.79^{\circ}$ in the extended zone scheme. (b) Lowest conduction band of the energy spectrum of bilayer graphene in the presence of an in-plane magnetic field with strength $B=10^4$ T. (c) and (d) show the location of the Dirac points of the energy spectrum in respectively (a) and (b) associated with the bottom (black) and top (red) layer. The different ellipses indicate pairs of Dirac points for which an equivalence between the two systems can be sought.}
\label{fig:commentplot1}
\end{figure}

The lowest conduction band of the energy spectrum governed by the Hamiltonian (\ref{eq:twisted_Hamiltonian}), is shown in Fig. \ref{fig:commentplot1}(a) in reciprocal space in the extended zone scheme\cite{kittel,bandfold} showing a region consisting of several Brillouin zones of the untwisted bilayer crystal. As a comparison, the lowest conduction band  of BLG with an in-plane magnetic field is shown in Fig. \ref{fig:commentplot1}(b). The figures show that one could approximate the energy spectrum around one of the original K-points of bilayer graphene, as is done in Ref. [\onlinecite{pseudo}], and obtain a similar Dirac point splitting as in the case of an in-plane magnetic field, as indicated by the blue full ellipses in Figs. \ref{fig:commentplot1}(c,d). However, if one would choose a different K-point in the first Brillouin zone, the splitting occurs in a different direction, as indicated by the orange dashed ellipses. This is not shared by the in-plane magnetic field model where the splitting occurs in the same direction for all Dirac points. One could ascribe this ambiguity in the direction of the pseudomagnetic field to the fact that pseudomagnetic fields are valley-dependent since they preserve time-reversal symmetry. However, choosing corners of the second or third Brillouin zone shows that the splitting becomes stronger, as indicated by the green dot-dashed ellipses, as well as being oriented in different directions. As such, there is also ambiguity in the strength of the pseudomagnetic field.

\begin{figure}[tb]
\centering
\includegraphics[width= 8.5cm]{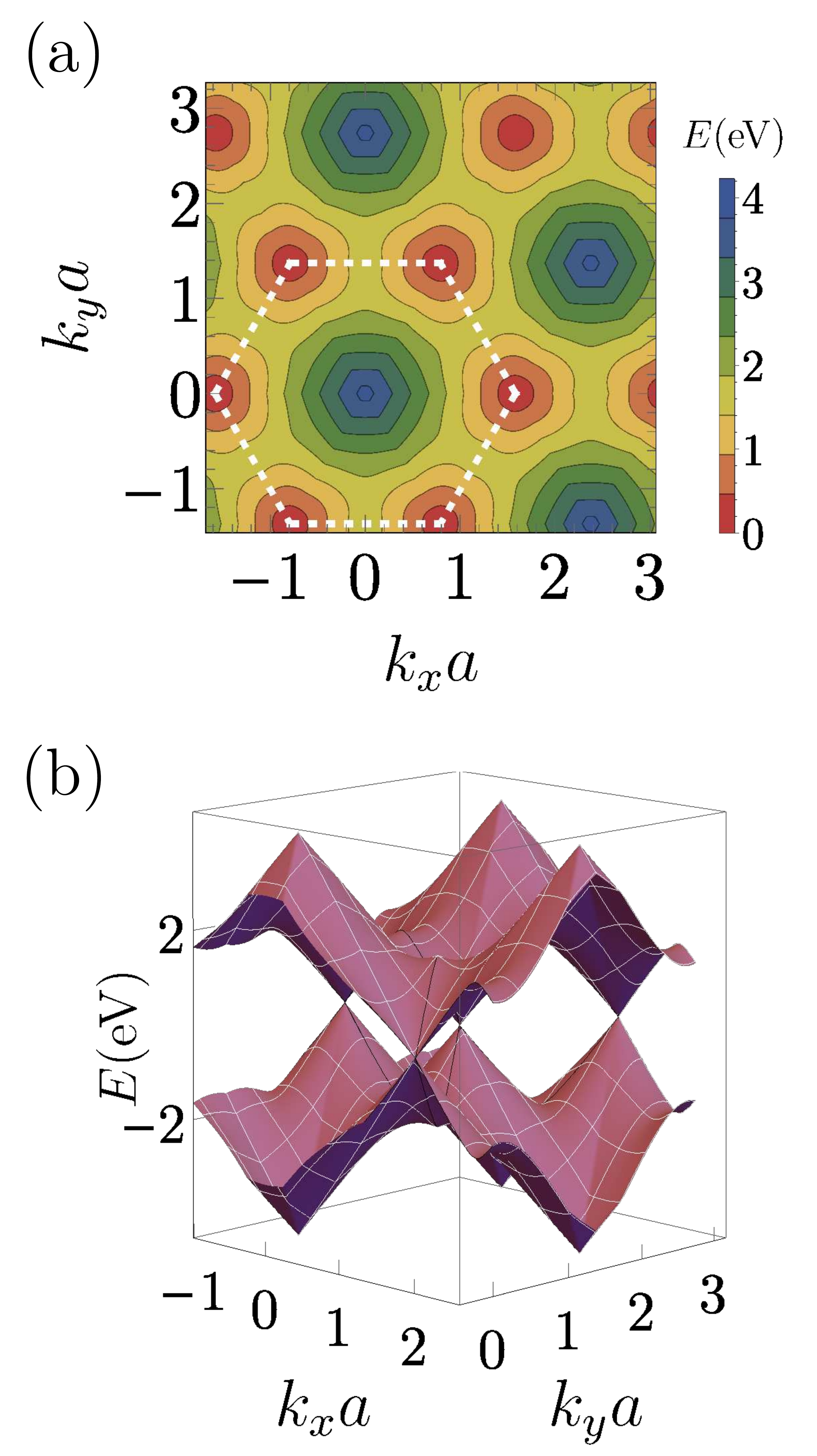}
\caption{(Color online) (a) Lowest conduction band of the energy spectrum of twisted bilayer graphene described by the Hamiltonian (\ref{eq:twisted_Hamiltonian}) for $\theta_t=21.79^{\circ}$ in the periodic zone scheme. The white dashed hexagon delimits the first sBZ. (b) 3D plot of the lowest conduction and valence band of the same system showing the linear Dirac cones which each are surrounded by three other cones.}
\label{fig:commentplot2}
\end{figure}

The arbitrariness in the choice of Dirac points that are considered in modeling electrons in twisted bilayer graphene can be resolved by examining the energy spectrum in the periodic zone scheme. As prescribed by standard band theory, the energy spectrum in the sBZ contains multiple bands in the reduced zone scheme that can be obtained by folding the energy spectrum of the Hamiltonian (\ref{eq:twisted_Hamiltonian}) back into the sBZ. The folding of the energy spectrum of the Hamiltonian (\ref{eq:twisted_Hamiltonian}) will map the monolayer-like Dirac cones onto each other. The zeroth order interlayer coupling in the considered Hamiltonians in Eq. (4) of Ref. [\onlinecite{pseudo}] and in Eq. (\ref{eq:twisted_Hamiltonian}) will, however, not couple these Dirac cones as it neglects coupling between regions connected by reciprocal superlattice vectors and will thus lead to doubly degenerate Dirac cones.

The periodic zone scheme is constructed by repeating the energy bands in the reduced zone scheme throughout the entire $\vec{k}$-space and is shown in Fig. \ref{fig:commentplot2}. The figures show that although the appearance of the Dirac cones in the low-energy spectrum is shared by both the in-plane magnetic field case and the twisting case described by the Hamiltonian (\ref{eq:twisted_Hamiltonian}), the two situations are not equivalent because in the case of twisted BLG the central Dirac point is accompanied by three other Dirac points\cite{contour1,contour2} while for BLG with an in-plane field, as seen in Fig. \ref{fig:commentplot1}, each Dirac point is accompanied by only one Dirac point. This clearly shows that it is not correct to relate these two phenomena.

The origin of the inequivalence between twisted BLG and BLG with an in-plane magnetic field, lies in the fact that they do not share the same spatial symmetry properties. The application of an in-plane magnetic field to bilayer graphene breaks $C_3$ symmetry and shifts all the Dirac points of one layer in the same direction. The Dirac points occur in pairs and the distance between them is proportional to the magnetic field. If the twisting angle is commensurate, twisted BLG preserves $C_3$ symmetry, which needs to be shared by the energy spectrum. While there is no $C_3$ symmetry around the Dirac points in the extended zone scheme in Figs. \ref{fig:commentplot1}(a) and (c), this symmetry is restored in the periodic zone scheme in Fig. \ref{fig:commentplot2}. Therefore, the Dirac points no longer occur in pairs and comparing this system to that of an in-plane magnetic field is not possible and no pseudomagnetic field is present. This is different from the case of perpendicular pseudomagnetic fields as a consequence of applying in-plane stress to a graphene flake\cite{pseudo2}, since a perpendicular magnetic field, as opposed to an in-plane magnetic field, preserves $C_3$ symmetry.

Furthermore, when considering states near the Dirac points, the coupling considered in the Hamiltonians in Eq. (4) of Ref. [\onlinecite{pseudo}] and in Eq. \eqref{eq:twisted_Hamiltonian} is a too crude approximation. The twisted bilayer graphene lattice leads to a spatial dependent interlayer coupling and recent works\cite{mele,mele2,koshino} have shown that this spatial dependence leads to coupling between the Dirac points in the periodic zone scheme. This effect is very important at low energies and leads to hybridization which can result in an AB stacked bilayer graphene spectrum or a gapped AA stacked bilayer graphene spectrum depending on the twist angle. In either case the low-energy spectrum is very different from a simple Dirac cone and as a consequence the low-energy transport properties will be very different as well, which is another argument why the equivalence between the effects of twisting and applying an in-plane magnetic field to bilayer graphene is invalid.

In conclusion, we have shown that there can be no in-plane pseudomagnetic fields generated by twisting bilayer graphene. This can be understood by means of simple symmetry arguments and is clear when studying the energy spectrum in the periodic zone scheme. In addition, the low-energy spectrum of twisted bilayer graphene is different from the monolayer-like Dirac cones of bilayer graphene in the presence of an in-plane magnetic field, implying that there will also be a difference between the transport properties of the charge carriers of these two systems.

\acknowledgments

This work was supported by the Research Foundation - Flanders (FWO-Vl) through aspirant research grants to MVDD and BVD.

\end{document}